\documentclass[12pt,draftcls,onecolumn]{IEEEtran}
\usepackage{amssymb,amsfonts,amsmath,stmaryrd,mathrsfs,verbatim,latexsym,epsfig,latexsym,enumerate,hyperref,footnote}
\usepackage[caption=false,font=footnotesize]{subfig}

\bstctlcite{IEEEexample:BSTcontrol}

\begin{document}

\title{Computational Intelligence Challenges and Applications on Large-Scale Astronomical Time Series Databases}

\author{Pablo Huijse, Millennium Institute of Astrophysics, Chile \\
Pablo A. Est\'evez*, Department of Electrical Engineering, Universidad de Chile and the Millennium Institute of Astrophysics, Chile \\
Pavlos Protopapas, School of Engineering and Applied Sciences, Harvard University, USA\\
Jos\'e C. Pr\'incipe, Computational NeuroEngineering Laboratory, University of Florida, USA \\
and Pablo Zegers, Facultad de Ingenier\'ia y Ciencias Aplicadas, Universidad de los Andes, Chile
\thanks{Copyright (c) 2013 IEEE. Personal use of this material is permitted. However, permission to use this material for any other purposes must be obtained from the IEEE by sending a request to pubs-permissions@ieee.org.}
\thanks{Pablo Huijse is with the Millennium Institute of Astrophysics, Chile and the Department of Electrical Engineering, Faculty of Physical and Mathematical Sciences, Universidad de Chile, Chile. Pablo Est\'evez* is with the Department of Electrical Engineering, Faculty of Physical and Mathematical Sciences, Universidad de Chile, Chile and the Millennium Institute of Astrophysics, Chile. *Pablo Est\'evez is the corresponding author.  Pavlos Protopapas is with the School of Engineering and Applied Sciences and the Center of Astrophysics, Harvard University, USA. Jos\'e Pr\'incipe is with the Computational Neuroengineering Laboratory, University of Florida, Gainesville, USA. Pablo Zegers is with the Universidad de los Andes, Facultad de Ingenier\'ia y Ciencias Aplicadas, Monse\~nor \'Alvaro del Portillo 12455, Las Condes, Santiago, Chile.  Correspondent email address: \textit{pestevez@ing.uchile.cl}.}}

\maketitle


\begin{abstract}
	Time-domain astronomy (TDA) is facing a paradigm shift caused by the exponential growth of the sample size, data complexity and data generation rates of new astronomical sky surveys. For example, the Large Synoptic Survey Telescope (LSST), which will begin operations in northern Chile in 2022, will generate a nearly 150 Petabyte imaging dataset of the southern hemisphere sky. The LSST will stream data at rates of 2 Terabytes per hour, effectively capturing an unprecedented movie of the sky. The LSST is expected not only to improve our understanding of time-varying astrophysical objects, but also to reveal a plethora of yet unknown faint and fast-varying phenomena. To cope with a change of paradigm to data-driven astronomy, the fields of astroinformatics and astrostatistics have been  created recently. The new data-oriented paradigms for astronomy combine statistics, data mining, knowledge discovery, machine learning and computational intelligence, in order to provide the automated and robust methods needed for the rapid detection and classification of known astrophysical objects as well as the unsupervised characterization of novel phenomena. In this article we present an overview of machine learning and computational intelligence applications to TDA. Future big data challenges and new lines of research in TDA, focusing on the LSST, are identified and discussed from the viewpoint of computational intelligence/machine learning. Interdisciplinary collaboration will be required to cope with the challenges posed by the deluge of astronomical data coming from the LSST.

\end{abstract}

\section{Introduction}


	Time domain astronomy (TDA) is the scientific field dedicated to the study of astronomical objects and associated phenomena that change through time, such as pulsating variable stars, cataclysmic and eruptive variables, asteroids, comets, quasi-stellar objects, eclipses, planetary transits and gravitational lensing,  to name just a few. The analysis of variable astronomical objects paves the way towards the understanding of astrophysical phenomena, and provides valuable insights in topics such as galaxy and stellar evolution, universe topology, and others.

	Recent advances in observing, storage, and processing technologies have facilitated the evolution of astronomical surveys from observations of small and focused areas of the sky (MACHO \cite{Alcock2000}, EROS \cite{EROS2009}, OGLE \cite{Udalski1997}) to deep and extended panoramic sky surveys (SDSS \cite{York2000}, Pan-STARRS \cite{Kaiser2002}, CRTS \cite{Catalina2003}). Data volume and generation rates are increasing exponentially, and instead of still images, future surveys will be able to capture digital ``movies of the sky'' from which variability will be characterized in ways never seen before.

	Several new grand telescopes are planned for the next decade \cite{Tyson2012}, among which is the Large Synoptic Survey Telescope (LSST \cite{LSST2012,LSST2013})  under construction in northern Chile and expected to begin operations by 2022. The word ``synoptic'' is used here in the sense of covering large areas of the sky repeatedly, searching for variable objects in position and time. The LSST will generate a 150 Petabyte imaging database, and a 40 Petabyte worth catalog associated with 50 billion astronomical objects during 10 years \cite{Borne2013}. The resolution, coverage, and cadence of the LSST will help us improve our understanding of known astrophysical objects and reveal a plethora of unknown faint and fast-varying phenomena \cite{LSST2013}. In addition, the LSST will issue approximately 2 million alerts nightly related to transient events, such as supernovae, for which facilities around the world can follow up.

	To produce science from this deluge of data the following open problems need to be solved \cite{Borne2012}: a) real-time mining of data streams of $\sim 2$ Terabytes per hour, b) real-time classification of the 50 billion followed objects, and c) the analysis, evaluation, and knowledge extraction of the 2 million nightly events.

    The big data era is bringing a change of paradigm in astronomy, in which scientific advances are becoming more and more data-driven \cite{Feilgelson2012}. Astronomers, statisticians, computer scientists and engineers have begun collaborations towards the solution of the previously mentioned problems, giving birth to the scientific fields of astrostatistics and astroinformatics \cite{Borne2010}. The development of fully-automated and robust methods for the rapid classification of what is known, and the characterization of emergent behavior in these massive astronomical databases are the main tasks of these new fields. We believe that computational intelligence, machine learning and statistics will play major roles in the development of these methods \cite{Feilgelson2012,Borne2013}.

	The remainder of this article is organized as follows: In section \ref{astrobg} the fundamental concepts related to time-domain astronomy are defined and described. In section \ref{review}  an overview of current computational intelligence (CI) and machine learning (ML) applications to TDA is presented. In section \ref{future} future big data challenges in TDA are exposed and discussed, focusing on what is needed for the particular case of the LSST from an ML/CI perspective. Finally, in section \ref{conclusion} conclusions are drawn.

\section{Astronomical Background} \label{astrobg}

	In this section we describe the basic concepts related to astronomical time series analysis and time-domain astronomical phenomena. Photometry is the branch of astronomy dedicated to the precise measurement of visible electromagnetic radiation from astronomical objects. To achieve this, several techniques and methods are applied to transform the raw data from the astronomical instruments into standard units of flux or intensity. The basic tool in the analysis of astronomical brightness variations is the \textbf{light curve}. A light curve is a plot of the magnitude of an object's electromagnetic radiation (in the visible spectrum) as a function of time.  
	
	Light curve analysis is challenging, not only because of the sheer size of the databases, but also due to the characteristics of the data itself. Astronomical time series are unevenly sampled due to constraints in the observation schedules, telescope allocations and other limitations. When observations are taken from Earth the resulting light curves will have periodic one-day gaps. The sampling is randomized because observations for each object happen at different times every night. 
	The cycles of the moon, bad weather conditions and sky visibility impose additional constraints which translate into data gaps of different lengths. Space observations are also restricted as they are regulated by the satellite orbits. Discontinuities in light curves can also be caused by technical factors: repositioning of the telescopes, calibration of equipment, electrical, and mechanical failures, etc.

	Astronomical time series are also affected by several noise sources. These noise sources can be broadly categorized into two classes. The first class is related to observations, such as the brightness of closer astronomical objects, and atmospheric noise due to refraction and extinction phenomena (scattering of light due to atmospheric dust). On the other hand, there are noise sources related to the  instrumentation, in particular to the CCD cameras, such as sensitivity variations of the detector, and thermal noise. In general, errors in astronomical time series are non-Gaussian and heterocesdastic, \emph{i.e.}, the variance of the error is not constant, and changes along the magnitude axis.
	
	Other common problematic situations arising in TDA are the sample-selection bias and the lack of balance between classes. Generally the astrophysical phenomena of interest represents a small fraction of the observable sky, hence the vast majority of the data belongs to the ``background class''. This is especially noticeable when the objective is to find unknown phenomena, a task known as novelty detection. Sufficient coverage and exhaustive labeling are required in order to have a good representation of the sample, and to assure capturing the rare objects of interests.
	
	\begin{figure}[t]
		  \centering
		  \includegraphics[scale=0.7]{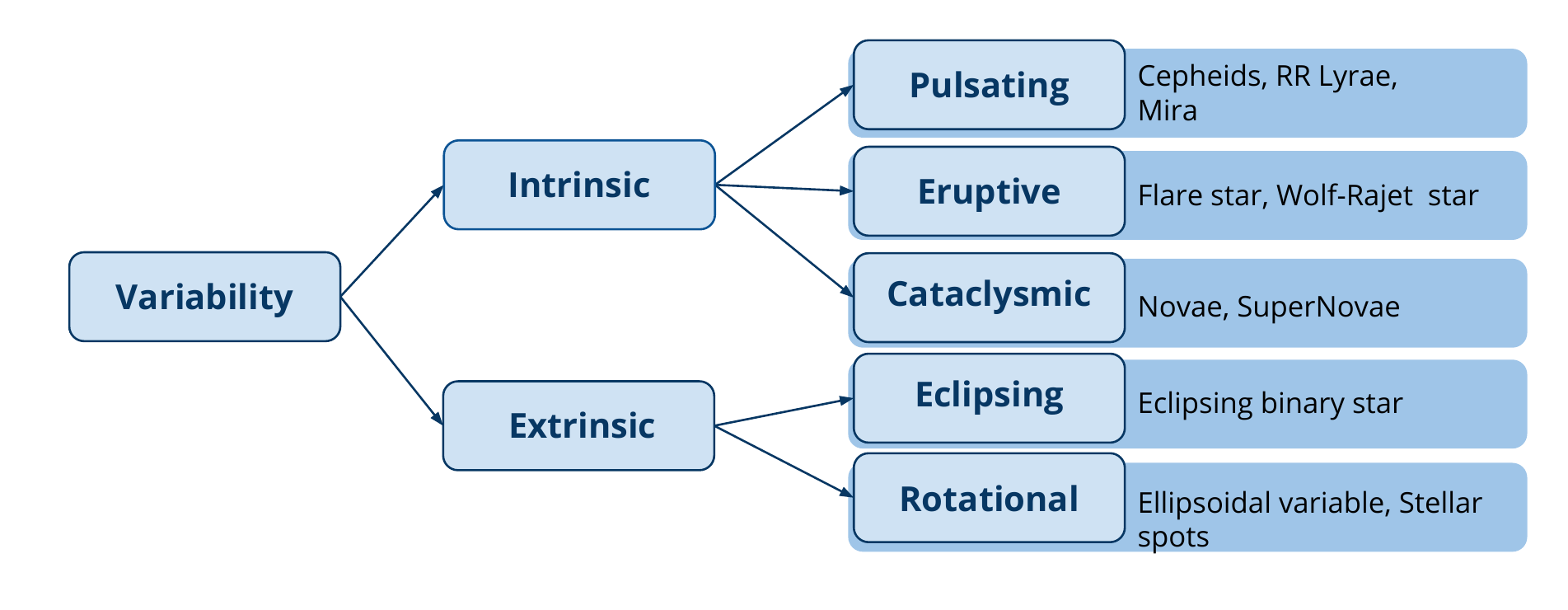}
		  \caption{\label{EyerClasses} Variable star topological classification. }
		  \vspace{-10pt}
	\end{figure}
	
	In the following we briefly describe several time-domain astronomical phenomena emphasizing their scientific interest. We focus on phenomena that vary in the optical spectrum. 
	Among the ``observable stars'' there is a particular group called the \textbf{variable stars} \cite{Petit1997,percy2007,Eyer2008}. Variable stars correspond to stellar objects whose brightness, as observed from Earth, fluctuates in time above a certain variability threshold defined by the sensitivity of the instruments. 
	Variable star analysis is a fundamental pivot in the study of stellar structure and properties, stellar evolution and the distribution and size of our Universe. The major categories of variable stars are briefly described in the following paragraphs with emphasis on the scientific interest behind each of them. For a more in-depth definition of the objects and their mechanisms of variability, the reader can refer to \cite{percy2007}. The relation between different classes of variable stars is summarized by the tree diagram shown in Fig. \ref{EyerClasses} \cite{Eyer2008,Samus2007}. 

\begin{figure}[t]
	  \centering
	  \subfloat[]{\label{pulsating} \includegraphics[scale=0.48]{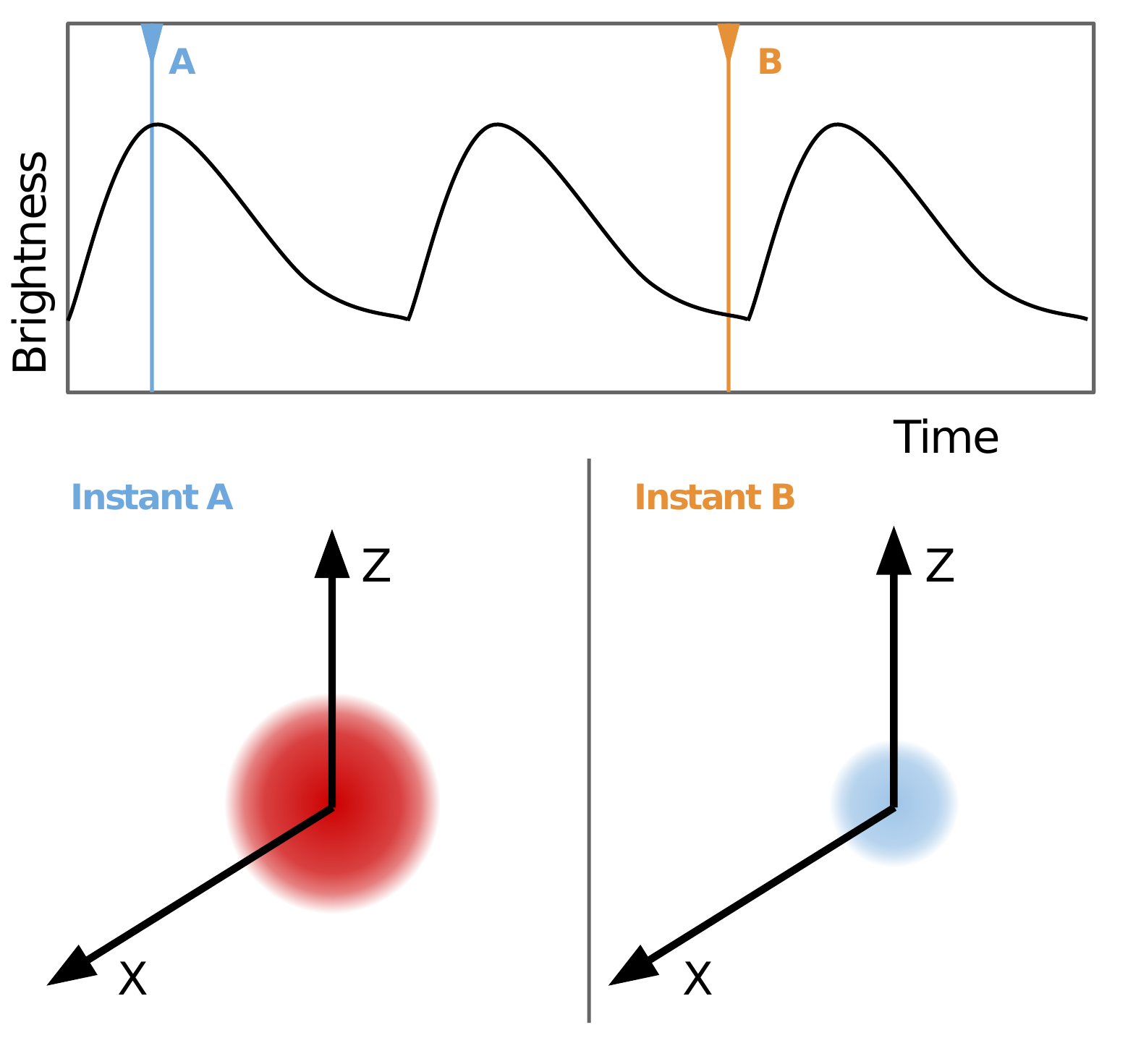}} 
	  \subfloat[]{\label{EB} \includegraphics[scale=0.48]{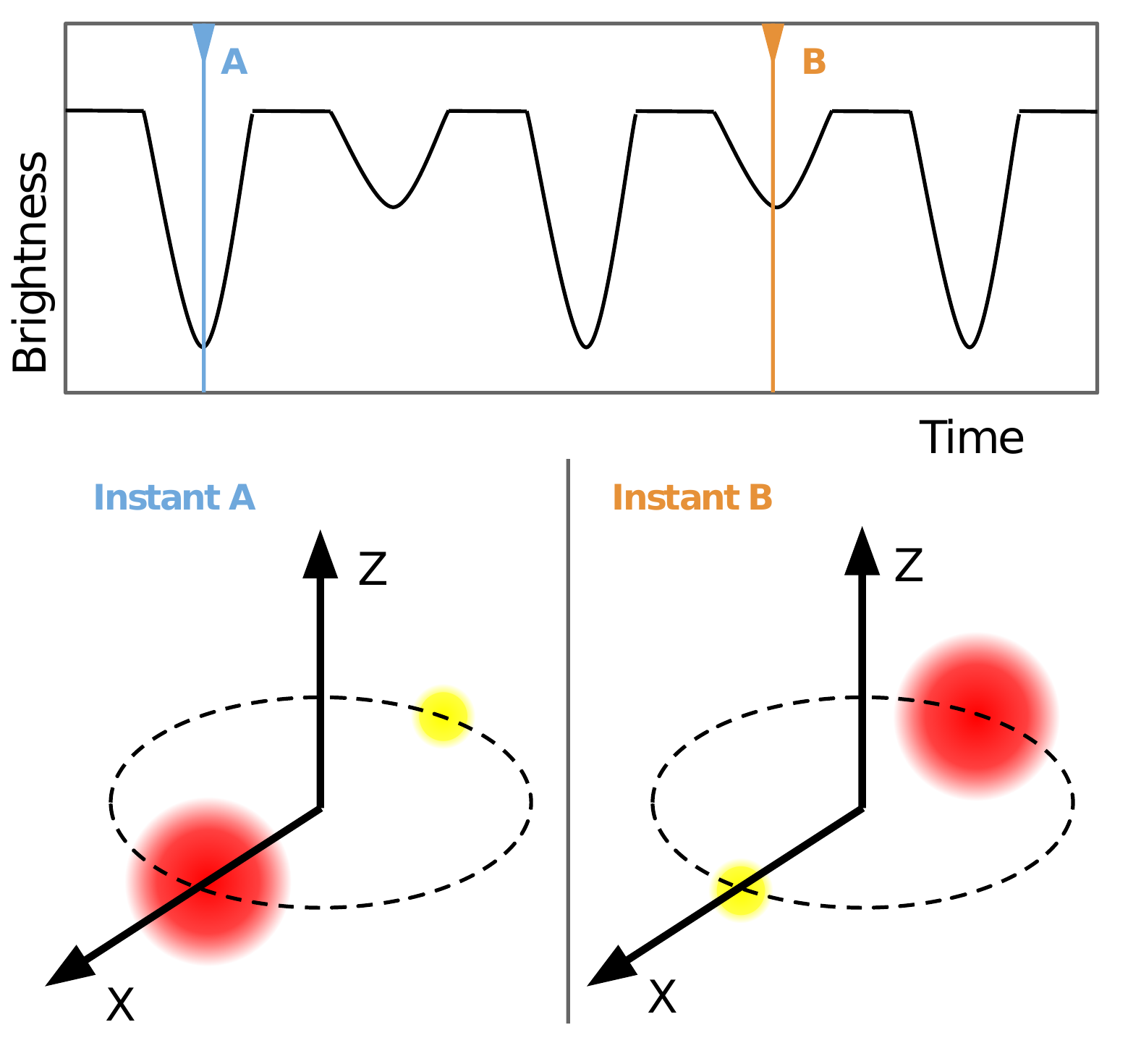}} 
	  
	  \caption{ (a) Light curve of a pulsating variable star (upper left panel), such as a Cepheid or RR Lyrae. The star pulsates periodically changing in size, temperature and brightness which is reflected on its light curve. (b) Light curve of eclipsing binary star (upper right panel). The lower panels show the geometry of the binary system at the instants where the eclipses occur. The periodic pattern in the light curve is observed because the Earth (X axis) is aligned with the orbital plane of the system (Z axis). }
		\vspace{-10pt}
		\end{figure}

\begin{figure}[t]
	  \centering
	  \subfloat[]{\label{RRLb} \includegraphics[scale=0.55]{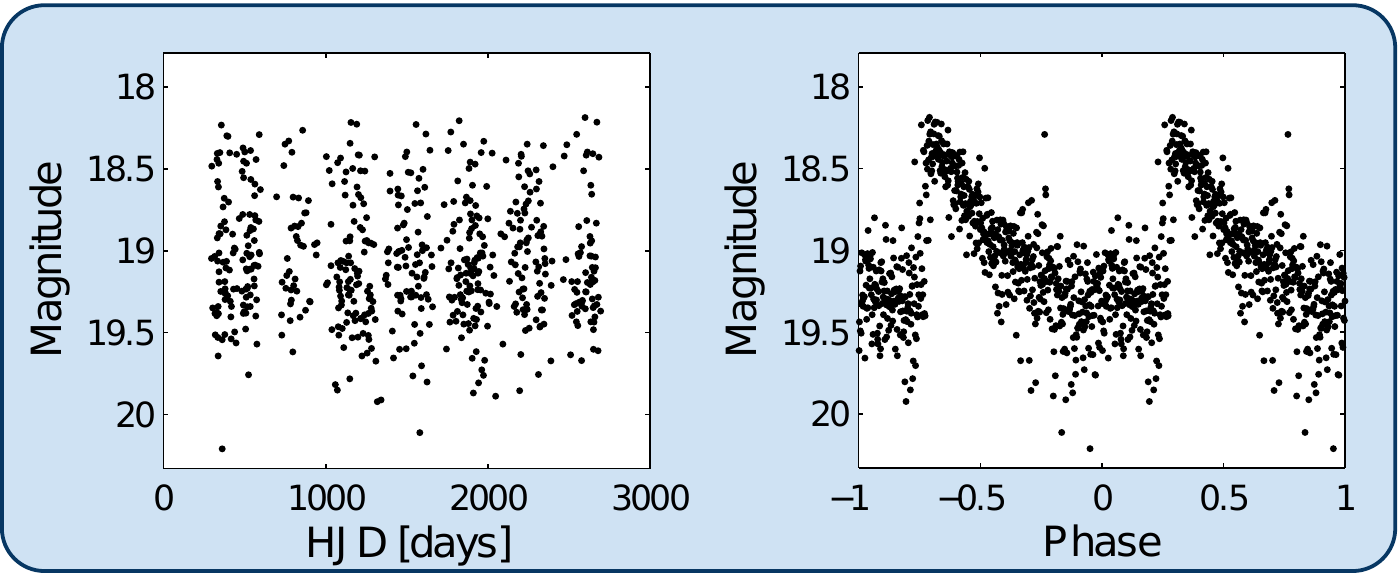}} 
	  \subfloat[]{\label{CEPHb} \includegraphics[scale=0.55]{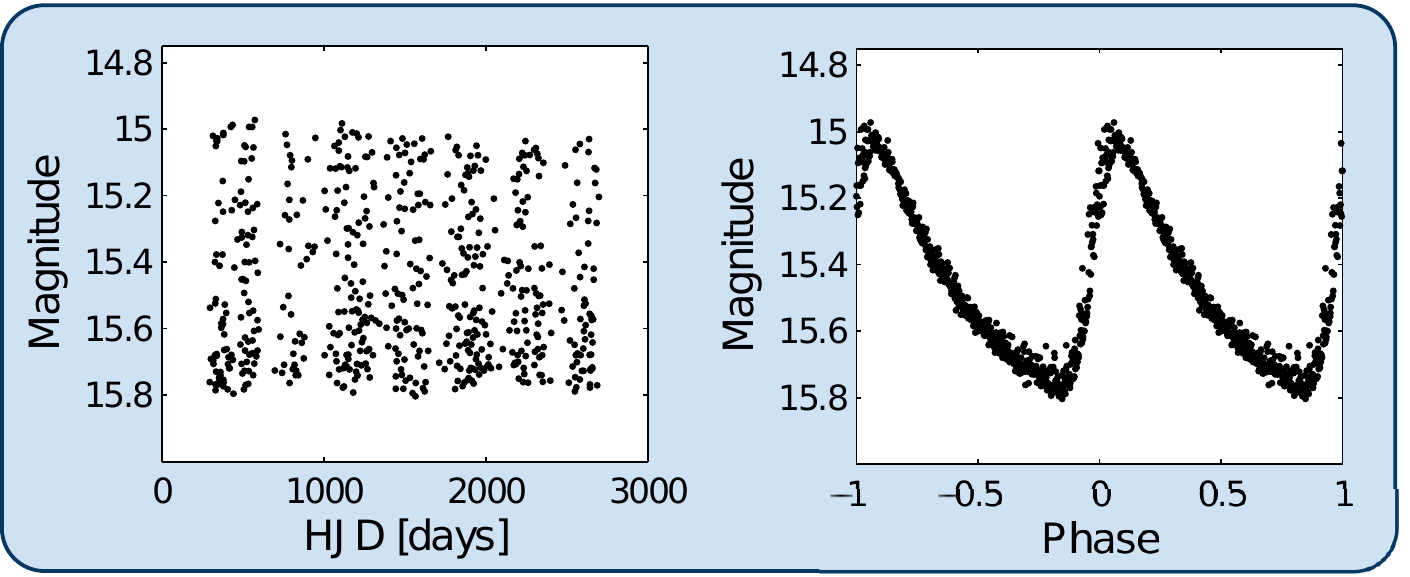}} 
	  \vspace{-10pt}
	  \subfloat[]{\label{MIRAb} \includegraphics[scale=0.55]{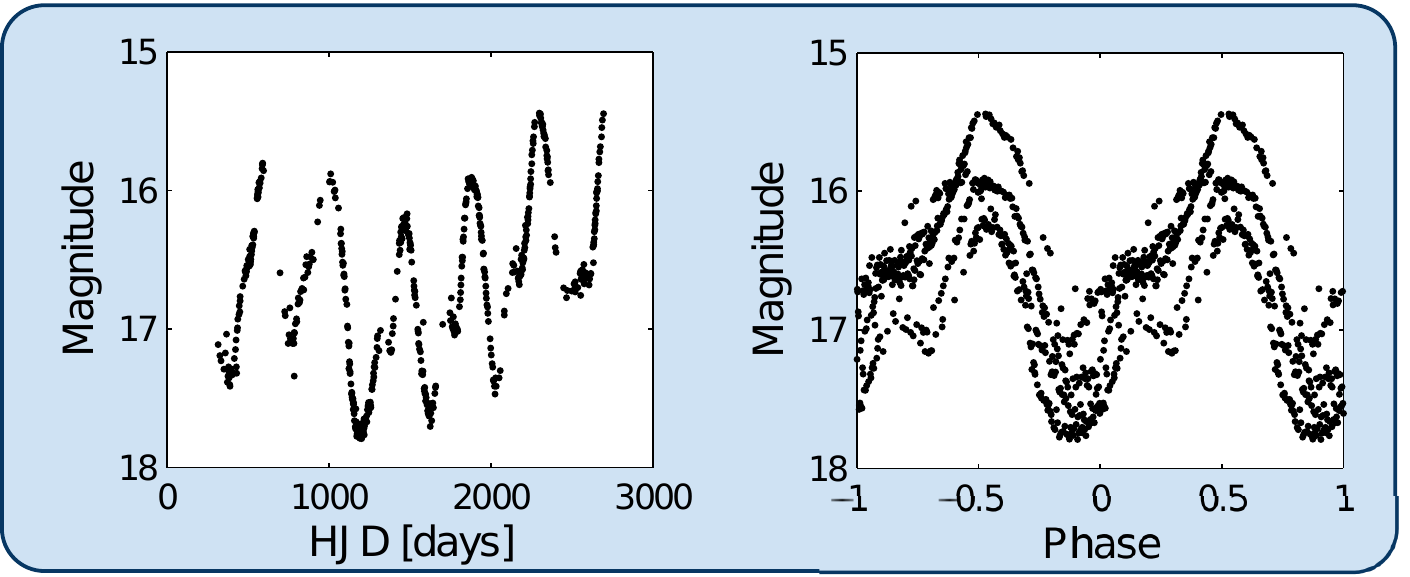}} 
	  \subfloat[]{\label{EBb} \includegraphics[scale=0.55]{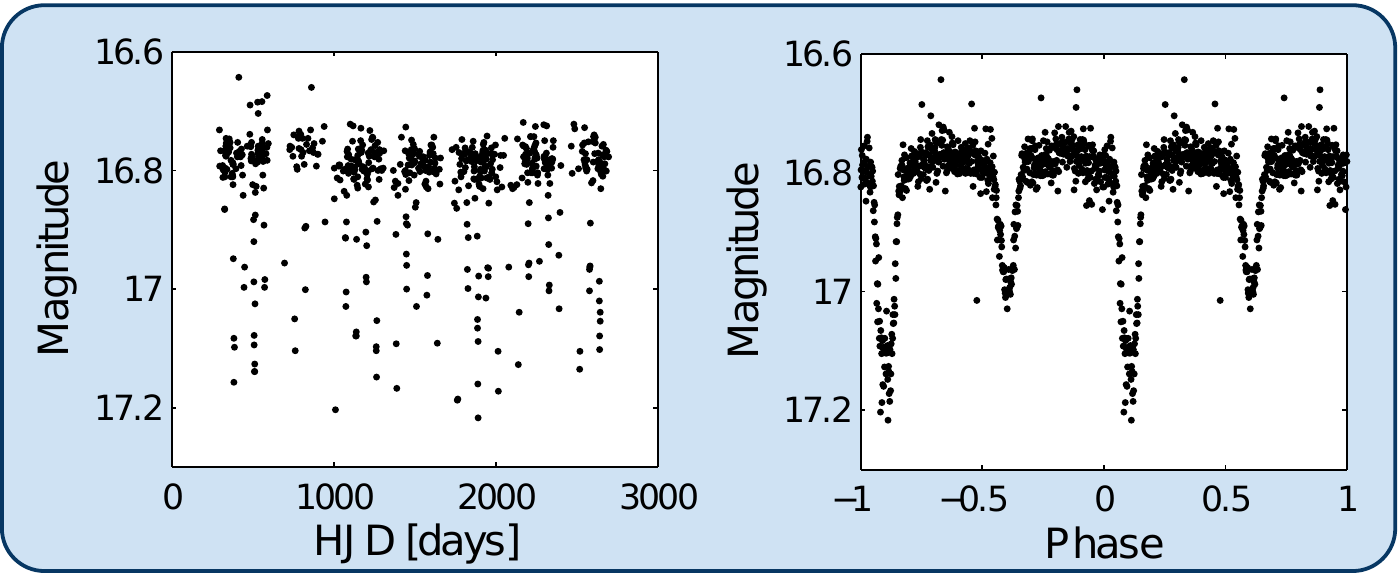}}

	  \caption{ \label{fig:var} Light curve and phase diagram of an RR Lyrae (a), Cepheid (b), Mira (c) and eclipsing binary star (d), respectively. The phase diagram is obtained using the underlying period of the light curves and the epoch folding  transformation \cite{percy2007}. If the folding period is correct a clear profile of the periodicity will appear in the phase diagram.} 
	\vspace{-10pt}
	\end{figure}

	
	The analysis of intrinsic variable stars is of great importance for the study of stellar nuclei and evolution. Some classes of intrinsic variable stars can  be used as distance markers to study the distribution and topology of the Universe. Cepheid and RR Lyrae stars \cite{percy2007} (Fig. \ref{pulsating}) are considered standard candles because of the relation between their pulsation period and their absolute brightness. It is possible to estimate the distance from these stars to Earth  with the period and the apparent brightness measured from the telescope \cite{Walker2003}. Type 1A Supernovae \cite{percy2007} are also standard candles, although they can be used to trace much longer distances than Cepheids and RR Lyrae \cite{Olivares2011}. 
	The period of eclipsing binary stars \cite{percy2007} (Fig. \ref{EB}) is a key parameter in astrophysics studies as it can be used to calculate the radii and masses of the components \cite{Popper1980}. 
	Light curves and phase diagrams of periodic variable stars are shown in Fig. \ref{fig:var}.

\section{Review of Computational Intelligence  Applications in TDA} \label{review}

	Time-domain astronomers are faced with a wide array of scientific questions that are related to the detection, identification and modeling of variable phenomena such as those presented in the previous Section. We may classify these problems broadly as follows:
	\begin{enumerate}[i]
		\item Extract information from the observed time series in order to understand the underlying processes of its source.
		\item Use previous knowledge of the time-varying universe to classify new variable sources automatically. How do we characterize what we know?
		\item Find structure in the data. Find what is odd and different from everything known. How do we compare astronomical objects? What similarity measure do we use?
	\end{enumerate}

	The computational intelligence and machine learning fields provide methods and techniques to deal with these problems in a robust and automated way. Problem i is a problem of modeling, parametrization and regression (kernel density estimation). Problem ii corresponds to supervised classification (artificial neural networks, random forests, support vector machines). Problem iii deals with unsupervised learning, feature space distances and clustering (\emph{k} nearest neighbors, self-organizing maps). The correct utilization of these methods is key to dealing with the deluge of available astronomical data. In the following section we review particular cases of computational intelligence based applications for TDA. 

\subsection{Periodic Variable Star Discrimination}


	We begin this review with a case of parameter estimation from light curves using information theoretic criteria. Precise period estimations are fundamental in the analysis of periodic variable stars and other periodic phenomena such as transiting exoplanets. 
	In \cite{Huijse2012} the correntropy kernelized periodogram (CKP), a metric for period discrimination for unevenly sampled time series, was presented. This periodogram is based on the correntropy function \cite{Principe2010}, an information theoretic functional that measures similarity over time using statistical information contained in the probability density function (pdf) of the samples. 
	In \cite{Huijse2012} the CKP was tested on a set of 5,000 light curves from the MACHO survey \cite{Alcock2000} previously classified by experts. The CKP achieved a true positive rate of 97\% having no false positives and outperformed conventional methods used in astronomy such as the Lomb-Scargle periodogram \cite{Scargle1982}, ANOVA and string length.

	In \cite{Protopapas2012} the CKP was used as the core of a periodicity discrimination pipeline for light curves from the EROS-2 survey \cite{EROS2009}. The method was calibrated using a set of 100,000 synthetic light curves generated from multivariate models constructed following the EROS-2 data. Periodicity thresholds and rules to adapt the kernel parameters of the CKP were obtained in the calibration phase. Approximately 32 million light curves from the Large and Small Magellanic clouds  were tested for periodicity. The pipeline was implemented for GPGPU architectures taking 18 hours to process the whole EROS-2 set on a cluster with 72 GPUs. A catalog of 120 thousand periodic variable stars was obtained and cross-matched with existing catalogs for the Magellanic clouds for validation. The main contributions of \cite{Protopapas2012} are the procedure used to create the training database using the available survey data, the fast implementation geared towards large astronomical databases, the large periodic light curve catalog generated from EROS-2, and the valuable inference on the percentage of periodic variable stars. 
	
	Another example of information theoretic concepts used for periodicity detection in light curves can be found in \cite{Grahamn2013MNRAS}. In this work the Shannon's conditional entropy of a light curve is computed from a binned phase diagram obtained for a given period candidate. The conditional entropy is minimized in order to find the period that produces the most ordered phase diagram. The proposed method was tested using a training set of periodic light curves from the MACHO survey and the results show that it is robust against systematic errors produced by the sampling, data gaps, aliasing and artifacts in phase space.  
	
	In Tagliaferri \emph{et al.} \cite{Tagliaferri1999}, neural networks are used to obtain the parameters of the periodogram of the light curve. These parameters are then fed into the MUSIC (Multiple Signal Classification) to generate a curve whose peaks are located in the periods sought. Interestingly, this work also shows the relation between the presented method and the Cramer-Rao bound, thus posing absolute practical limits to the performance of the proposed procedure.
	
	A comprehensive analysis of period finding algorithms for light curves can be found in \cite{Graham2013MNRAS2}. In this work classical methods for period discrimination in astronomy such as the Lomb-Scargle periodogram and Phase Dispersion minimization are compared to novel information theoretic criteria \cite{Grahamn2013MNRAS}. The authors note that  the accuracy of each individual method is dependent on observational factors and suggest that an ensemble approach that combines several algorithms could mitigate this effect and provide a more consistent solution. How to combine the output of different methods and the increased  computational complexity are key issues to be solved.

\subsection{Automated Supervised Classification for Variable Objects}

	After obtaining the period, supervised methods can be used to discriminate among the known classes of periodic variable stars. In supervised learning, prior information in the form of a training dataset is needed to classify new samples. The creation and validation of these training sets are complex tasks in which human intervention is usually inevitable. This is  particularly challenging in the astronomical case due to the vast amounts of available data. If the data do not initially represent the population well, then scientific discovery may be hindered. In addition, due to the differences between observational surveys, it is very difficult to reuse the training sets. In the following paragraphs several attempts of supervised classification for TDA are reviewed, with emphasis on the classification scheme and the design of the training databases.

\begin{figure}[t]
	  \centering
	  \subfloat[]{\label{dclass} \includegraphics[scale=0.55]{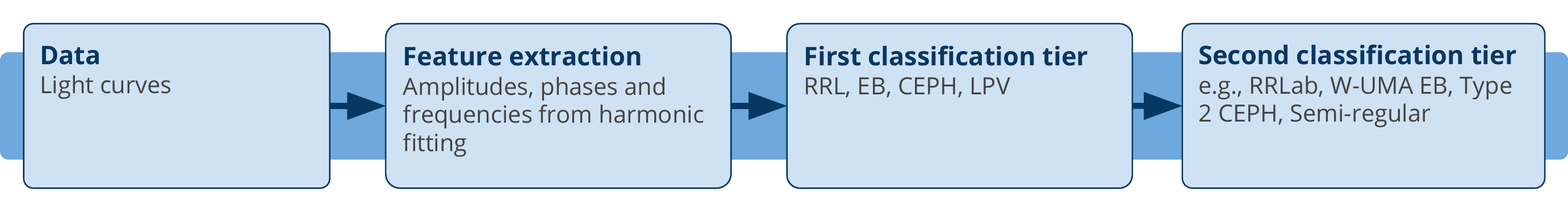}} \\ 
	  \subfloat[]{\label{pclass} \includegraphics[scale=0.55]{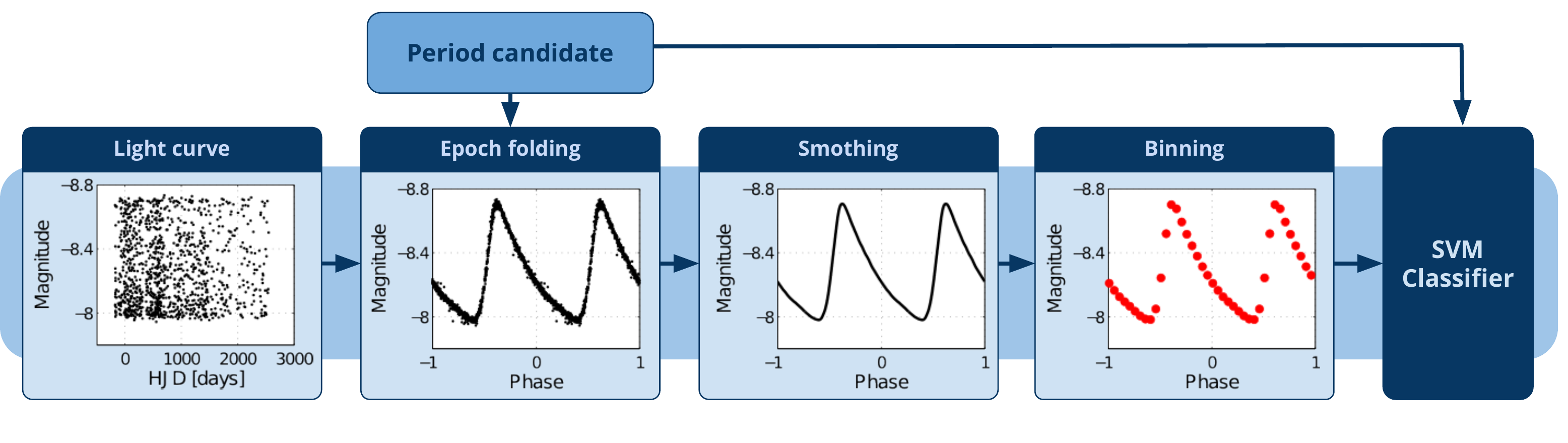}} 
	  
	  \caption{ (a) Classification scheme used in \cite{Debosscher2007} for variable star classification. (b) Light curve processing pipeline used in \cite{Wachman2009}. A candidate period is used to obtain a phase diagram of the light curve which is then smoothed, interpolated to a fixed grid, and binned. The period, magnitude, color and binned values are used as features for the SVM classifier. }
	  \vspace{-10pt}
	\end{figure}

	Gaussian mixture models (GMMs), and artificial neural networks trained through Bayesian model averaging (BAANN) were used to discriminate periodic variable stars from their light curves in \cite{Debosscher2007}. Classification schemes using single multi-class and multi-stage (hierarchical) classifiers were compared. This work is relevant not only because of the application and comparison between the algorithms, but also because of the extended analysis performed in building the training dataset. First, well-known class prototypes were recovered from published catalogs and reviewed. Data from nine astronomical surveys were used, although the vast majority came from the Hipparcos and OGLE projects. A diagram of the classification pipeline for the hierarchical classifier is shown in Fig. \ref{dclass}. The selected variability types were parametrized using harmonic fitting (Lomb-Scargle periodogram). The classes were organized in a tree-like structure similar to the one shown in Fig. \ref{EyerClasses}. The final training set contained 1,732 samples from 25 well-represented classes. For the single stage classifier the GMM and BAANN obtained correct classification rates of 69\% and 70\%, respectively. Only the BAANN was tested using the multi-stage scheme obtaining a correct classification rate of 71\%. According to the authors, the GMM provides a simple solution with direct astrophysical interpretation. On the other hand, some machine learning algorithms may achieve lower misclassification rates but their interpretability is reduced. The authors also state the need for higher statistical knowledge, which can be provided through interdisciplinary cooperation, in order to use the machine learning approach. A more recent version of this method can be found in \cite{Blomme2011}. In this work 26,000 light curves from TrES and Kepler surveys were classified using a multi-stage tree of GMMs. The main difference from the previous version is the careful selection of significant frequencies and overtone features which reduces confusion between classes.


	In \cite{Wachman2009}, 14,087 periodic light curves from the OGLE survey were used to train and test supervised classifiers based on \emph{k}-NN and SVM. The periods and labels were obtained directly from the OGLE survey. The following periodic variable classes were considered: Cepheids, RR Lyrae and Eclipsing Binaries. The period, average brightness and color of the light curves were used as features for the classifier.  In addition, the authors included the phase diagram of the light curve as a feature and proposed a kernel to compare time series, which is plugged into the SVM. The phase diagram is obtained using the underlying period of the light curves and the epoch folding  transformation \cite{percy2007}. The light curve processing pipeline is shown in Fig. \ref{pclass}. The proposed kernel takes care of the possible difference in phase between time series. Using the shape of the light curve and the proposed kernel, correct classification rates close to 99\% were obtained. Intuitively, the shape of the periodicity (phase diagram) should be a strong feature for periodic variable star classification. The authors note that a complete pipeline would require first discriminating whether the light curve is periodic, and estimating its period with high accuracy. Wrongly estimated periods would alter the folded light curve, affecting the classification performance. 
	
	A Random Forest (RF) classifier for periodic variable stars from the Hipparcos survey was presented in \cite{Dubath2011}. A set of 2,000 reliable variable sources found in the literature was used to train the classifier to discriminate 26 types of periodic variable stars. Non-periodic variables were also added to the training set. Light curves were characterized using statistical moments, periods and Fourier coefficients.  The performance of the classifier is consistent with other studies \cite{Debosscher2007}. The authors found that the most relevant features in decreasing order are the period, amplitude, color and the Fourier coefficients (light curve model). The authors also found that the major sources of misclassification are related to the reliability of the estimated periods and the misidentification of non-periodic light curves as periodic.
	
	Statistical classifiers work under the assumption that class probabilities are equivalent for the training and testing sets. According to \cite{Long2012} this assumption may not hold when the number of light curve measurements is different between sets. In \cite{Long2012} this problem is addressed via noisification and denoisification, \emph{i.e.}, trying to modify the pdf of the training set so that it mimics the test set, and to infer the class of a poorly sampled time series according to its most probable evolution. This scheme is tested on light curves from the OGLE survey. Results show that noisification and denoisification improve the classification accuracy for poorly sampled time series by 20\%. The authors note that the proposed method may help overcome other systematic differences between sets such as varying cadences and noise properties.

	Classification of non-periodic variable sources is less developed than periodic source classification. Non-periodic sources in general are more difficult to characterize, which is why only a few studies do general Active Galactic Nuclei (AGN) classification, instead focusing on discriminating a particular type of quasi-stellar object (QSO). In \cite{Kim2011} a supervised classification scheme based on SVM was used to discriminate AGN from their light curves. The objects were characterized using 11 features including amplitude, color, autocorrelation function, variability index and period. A training set of $\sim$ 5,000 objects including non-variable and variable stars and 58 known quasars from the MACHO survey was used to train the classifier. This work \cite{Kim2011} differs from previous attempts at AGN discrimination in its thoughtful study of the efficiency and false positive rates of the model and classifier. The classifier was tested on the full 40 million MACHO light curves finding 1,620 QSO candidates that were later cross-matched with external quasar catalogs, confirming the validity of the candidates.

	In \cite{Richards2011}, a multi-class classifier for variable stars was proposed and tested. This implementation shares the features, data, and the 25-class taxonomy used in \cite{Debosscher2007}, allowing direct comparison. An important distinction with respect to \cite{Debosscher2007} is that the classification is extended to include non-periodic eruptive variables\footnote{Simple statistical moments from the flux distribution are used as features for the eruptive variables.}. Two classification schemes are tested: In the first, pairwise comparisons between two-class classifiers are used, and in the second, a hierarchical classification scheme based on the known taxonomy of the variable stars is employed (Fig. \ref{EyerClasses}). The best performance is obtained by the RF with pairwise class comparisons, achieving a correct classification rate of 73.3\% when using the same features as \cite{Debosscher2007}, and 77.2\% when only the more relevant features are used. In a taxonomical sense, a mistake committed in the first tier of the classification hierarchy (catastrophic error) is more severe than a mistake in the final tiers (sub-type classifiers). The hierarchical RF implementation obtains a slightly worse overall performance (1\%) and a smaller catastrophic error (8\%) than the pairwise RF. Although there is a considerable improvement with respect to \cite{Debosscher2007}, the accuracy is not high enough for fully automated classification. The taxonomical classification of variable stars and their multiple sub-types is still an open problem.

	Using information from several catalogs may improve the characterization of the objects under study, but joining catalogs is not a trivial task as different surveys use different instruments and may be installed in totally different locations. Intersecting the catalogs, \emph{i.e.}, removing columns/rows with missing data may result in a database that is smaller than the original single catalogs. A variable star classifier for incomplete catalogs was proposed in \cite{Pichara2013}. In this work the structure of a Bayesian network is learned from a joined catalog with missing data from several surveys. The joined catalog is ``filled'' by estimating the probability distributions and dependencies between the features through the Bayesian network. The resulting training set has 1,833 samples including non-variables, non-periodic variables (quasars and Be stars), and periodic variable stars (Cepheids, RR Lyrae, EB and Long Period Variables). An RF classifier is trained with the joined catalog. The Bayesian network is compared to a second approach for filling missing data based on GMM, obtaining better classification accuracy for most of the selected classes. An additional test on 20 million light curves from the MACHO survey (a catalog with no missing data) was performed. From this test a set of 1,730 quasar candidates was obtained which corresponds to a 15\% improvement with respect to previous quasar lists in the literature \cite{Kim2011}. 

\subsection{Unsupervised and Semi-supervised Learning in Time-domain Astronomy}

	In some cases previous information on the phenomena might be insufficient or non-existent, hence a training set cannot be constructed. In these cases one may need to go back one step and obtain this information from the data using unsupervised methods. In a broad sense the objective of unsupervised methods is to estimate the density function that generated the data revealing its structure. One of the first references for unsupervised learning in astronomy is found in \cite{Hernandez1994} where self-organizing maps (SOMs) were used to discriminate clusters of stars in the solar neighborhood. One hundred thousand stars from the Hipparcos catalog, mixed with synthetic data, were used. The synthetic stars were modeled with particular characteristics of known stellar populations. The Hipparcos catalog provides information about the position, magnitude (brightness), color\footnote{The color corresponds to the difference in average brightness between two different spectra.}, spectral type and variability among other features. The SOM was trained on a 10x10 grid with the additional constraint that each node should have at least one synthetic star. Using the synthetic stars as markers, clusters of stellar populations were recognized in the visualization of the SOM. In this case the SOM was used not only to find population clusters but also to validate the theoretical models used to create the synthetic stars.

	SOM was also used in \cite{Brett2004} in order to learn clusters of mono-periodic variable stars. The main objective was to classify periodic variable stars in the absence of a training set, which was the reason the SOM was selected. The feature in this case was an $N$-dimensional vector obtained from the folded light curves. Each light curve was folded with its period which was known a priori. The folded light curves were then normalized in scale and discretized in $N$ bins. The number of bins and SOM topology parameters were calibrated using five thousand synthetic light curves representing four classes of periodic variables. The SOM was then tested on a relatively small set of 1,206 light curves. The clusters for each class were discriminated using a U-matrix visualization and density estimations. Clusters associated with Eclipsing Binaries, Cepheids, RR Lyrae and $\delta$-scuti populations were identified. Although restricted in size, this application shows the potential of the SOM for class discovery in time-domain astronomy.
	
	In \cite{Sarro2009} a density based-approach for clustering was used to find groups of variable stars in the OGLE and CoRoT surveys. The light curves are characterized using the features described in \cite{Debosscher2007}. Each point in feature space is assigned to one of the clusters using Modal Expectation Maximization. This work addresses the need of tying up astronomical knowledge with the outcome of the computational intelligence algorithms. The manner in which astronomers have classified stellar objects and events does not necessarily correspond to that produced by automated systems. Interestingly, this study establishes that there is another problem as well: the same computational intelligence algorithms working on different databases produced distinct classification structures, showing that even though these databases have large numbers of examples, they have inherent biases and may not be sufficiently large to allow the discovery of general rules. This problem has also been reported in other fields, specifically artificial vision \cite{Torralba2008}. The work in \cite{Torralba2008} showed that in order to produce consistent classification performances, one could not simply use databases with hundreds of thousands of examples, it was necessary to use close to 80 million images, far exceeding what was traditionally considered enough by the practitioners of the field.
	
	Kernel Principal Component Analysis (KPCA) was used in \cite{Varon2011} to perform spectral clustering on light curves from the CoRoT survey. The light curves were characterized using three different approaches: Fourier series, autocorrelation functions, and Hidden Markov Models (HMMs). Then, dimensionality was reduced with KPCA using the Gaussian kernel. Finally, the eigenvalues were used to find clusters of variable stars. This novel characterization of light curves permits  identifying not only periodic variable stars correctly (Fourier and autocorrelation features), but also irregular variable stars (HMM features).

	Unsupervised learning can also be used for novelty detection, \emph{i.e.}, finding objects that are statistically different from everything that is known and hence cannot be classified in one of the existing categories. Astronomy has a long history regarding serendipitous discovery \cite{Fabian2009}, \emph{i.e.}, to find the unexpected (and unsought). 
	Computational intelligence and machine learning may provide the means for facilitating the task of novelty detection.

	One may argue that the first step for novelty detection is to define a similarity metric for astronomical time series in order to compare time-varying astronomical objects. This is the approach found in \cite{Protopapas2006} where a methodology for outlier light curve identification in astronomical catalogs was presented. A similarity metric based on the correlation coefficient is computed between every pair of light curves in order to obtain a similarity matrix. Intuitively, the outlier variable star will be dissimilar to all the other variables. Before any distance calculation, light curves are interpolated and smoothed in order to normalize the number of points and time instants. For each pair the lag of maximum cross-correlation is found in Fourier space, which solves the problem of comparison between light curves with arbitrary phases. Finally the outliers correspond to the light curves with the lowest cross correlations with respect to each row of the distance matrix. The method was tested on $\sim$ 34,500 light curves from early-stage periodic variable star catalogs originated from the MACHO and OGLE \cite{Udalski1997} surveys. The results of this process were lists of mislabeled variables with careful explanations of the new phenomena and reasons why they were misclassified. 
	Calculating the similarity matrix scales quadratically with the number of light curves in the survey. The authors discuss this issue and provide an approximation of the metric that reduces the computational complexity to  ${\cal O}(N)$. Also, an exact and efficient solution for distance-based outlier detection can be found in \cite{Dragomir2008}, which uses a discord metric that requires only two linear searches to find outlier light curves as well.


	A different approach for novelty detection was given in \cite{Rebbapragada2009}, where an anomaly detection technique dubbed PCAD (Periodic Curve Anomaly Detection) was proposed and used to find outlier periodic variables in large astronomical databases. PCAD finds clusters using a modified \emph{k}-means algorithm called phased \emph{k}-means (\emph{pk}-means). This modification is required in order to compare asynchronous time series (arbitrary phases). By using a clustering methodology the authors were able to find anomalies in both a global and local sense. Local anomalies correspond to periodic variables that lie in the frontier of a given class. Global anomalies on the other hand differ from all the clusters. Approximately 10,000 periodic light curves from the OGLE survey were tested with PCAD. The pre-processing of the light curves, the selection of features, and the computation of the cross-correlations follow the work of \cite{Protopapas2006}. The cross-correlation is used as a distance metric for the \emph{pk}-means. The results obtained by the method were then evaluated by experts and sorted as noisy light curves, misclassified light curves and interesting outliers worthy of follow-up.
	
\begin{figure}[t]
	  \centering
	  \subfloat[]{\label{SS} \includegraphics[scale=0.55]{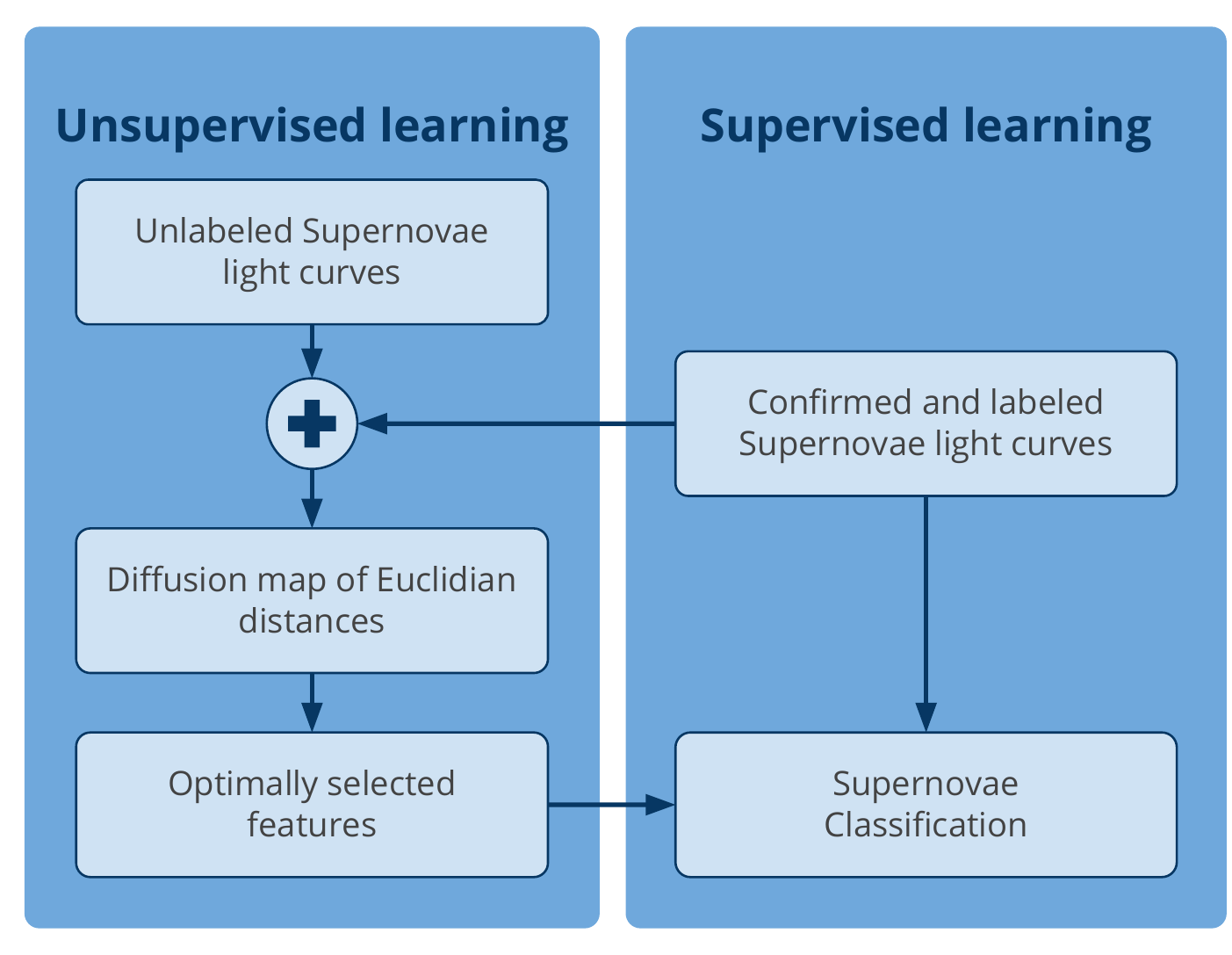}} 
	  \subfloat[]{\label{AL} \includegraphics[scale=0.55]{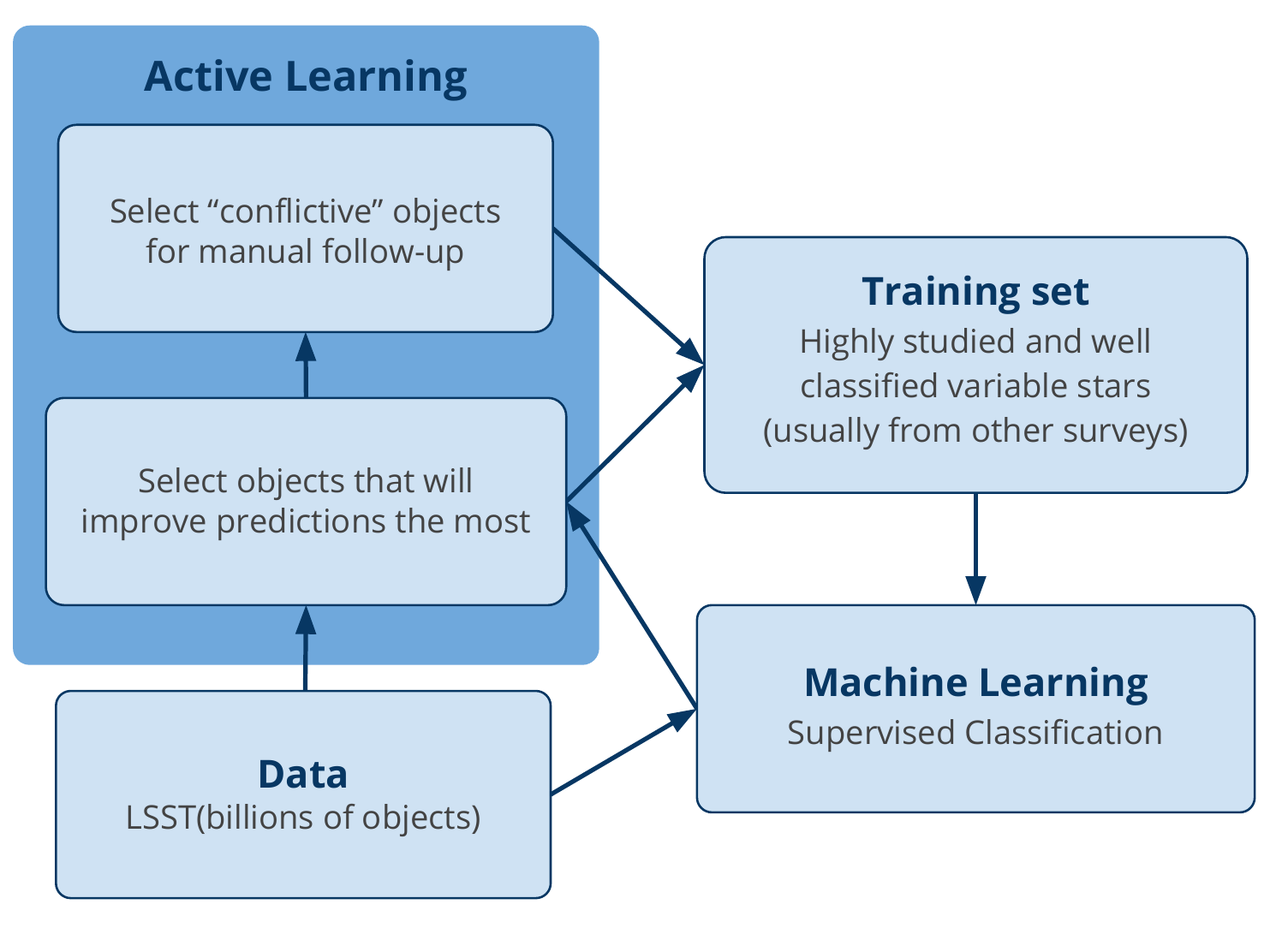}} 
	  
	  \caption{ (a) Semi-supervised learning scheme used in \cite{Richards2012-2} for Supernovae classification. (b) An active learning approach to building training sets for supervised classification of variable stars. Samples from the testing set are moved (unsupervised) to the training sets reducing sample selection bias. The expert is queried by the method if more information is needed for obtaining the new labels. }
	  \vspace{-10pt}
	\end{figure}

	A problem with purely unsupervised methods is that prior knowledge, when available, is not necessarily used. Semi-supervised learning schemes deal with the case where labels (supervised information) exist although not for all the available data. Semi-supervised methods are able to find the structure of the data distribution, learn representations and then combine this information with what is known. Semi-supervised methods can also be used for novelty detection, with the benefit that they may improve their discrimination by automatically incorporating the newly extracted knowledge. The semi-supervised approach is particularly interesting in the astronomical case where prior information exists, although scarce in comparison to the bulk of available unlabeled data. In \cite{Richards2012-2} a semi-supervised scheme for classification of supernova sub-types  was proposed. In this work the unlabeled supernovae data are used to obtain optimal low-dimensional representations in an unsupervised way. A diagram of the proposed implementation is shown in Fig. \ref{SS}. In general, features are extracted from supernovae light curves following fixed templates. The data-driven feature extraction proposed in \cite{Richards2012-2} performs better and is more efficient than template methods with respect to data utilization and scaling.
	
	A bias due to sample selection occurs when training and test datasets are not drawn from the same distribution. In astronomical applications, training datasets often come from older catalogs which, because of technological constraints, contain more information on brighter and closer astronomical objects, \emph{i.e.}, the training dataset is a biased representation of the whole. In these cases standard cross-validation procedures are also biased resulting in poor model selection and sub-optimal prediction. The selection bias problem is addressed from an astronomical perspective in \cite{Richards2012} through the use of active learning (AL). A diagram of the implementation proposed in \cite{Richards2012} is shown in Fig. \ref{AL}. In AL, the method queries the expert for manual follow-up of objects that cannot be labeled automatically. There is a natural synergy between AL and astronomy, because the astronomer is, in general, able to follow up a certain target in order to obtain additional information. The AL classifier consistently performed better than traditional implementations.

\section{Future Big Data Challenges in Time Domain Astronomy} \label{future}

	In this section we describe the future big data challenges in TDA from the viewpoint of computational intelligence and machine learning using as an example the LSST. The astronomical research problems targeted by the LSST are described in the LSST Science Book \cite{LSST2013}.  The LSST focuses on time-domain astronomy, and will be able to provide a movie of the entire sky in the southern hemisphere for the first time. Some of the LSST challenges are:  detecting faint transient signals such as supernovae with a low false-positive rate, classifying transients, estimating the distance from Earth, and making discoveries and classification in real-time \cite{Feilgelson2012}. Facilities such as the LSST will produce a paradigm change in astronomy. On the one hand the new telescope will be entirely dedicated to a large-scale survey of the sky, and individual astronomers will not be allowed to make private observations as they used to do in the past \cite{Feilgelson2012}.  On the other hand, the data volume and its flow rate would be so large that most of the process should be done automatically using robotic telescopes and automated data analysis.  Data volumes from large sky surveys will grow from Terabytes during this decade (\emph{e.g.}, PanSTARRS \cite{Kaiser2002}) to hundreds of Petabytes during the next decade (LSST \cite{LSST2012,LSST2013}). The final LSST image archive will be $\sim$ 150 Petabytes  and the astronomical object catalog (object-attribute database) is expected to be $\sim$ 40 Petabytes, comprising 200 attributes for 50 billion objects \cite{Borne2013}.  In \cite{Borne2012} the following three challenges are identified for the LSST: a) Mining a massive data stream of $\sim$ 2 Terabytes per hour in real time for 10 years, b) classifying more than 50 billion objects and following up many of these events in real time, c) extracting knowledge in real time for $\sim$ 2 million events per night. 
	The analysis of astronomical data involves many stages, and in all of them it is possible to use CI techniques to help its automation. In this paper we focused on the analysis of the light curves, but there are other CI challenges in image acquisition \& processing \cite{LSST2013}, dimensionality reduction and feature selection \cite{Donalek2013,Vergara2014}, etc. There are also major technical challenges related to the astronomical instruments, storage, networking and processing facilities. The challenges associated with data management and the proposed engineering solutions are described in \cite{LSST2012}.   

	The LSST will be dedicated exclusively to the survey program, thus follow-up observations (light curves, spectroscopy, multiple wavelengths), which are scientifically essential, must be done by other facilities around the world \cite{Borne2013}.  With this goal the LSST will generate millions of event alerts during each night for 10 years. Many of the observed phenomena are transient events such as supernovae, gamma-ray bursts, gravitational microlensing events, planetary occultations, stellar flares, accretion flares from supermassive black holes, asteroids, etc. \cite{Djorgovski2006}. A key challenge is that the data need to be processed as it streams from the telescopes, comparing it with the previous images of the same parts of the sky, automatically detecting any changes, and classifying and prioritizing the detected events for rapid follow-up observations \cite{Graham2012}. The system should output a probability of any given event as belonging to any of the possible known classes, or as being unknown. An important requirement is maintaining high level of completeness (do not miss any interesting events) with a low false alarm rate, and the capacity to learn from past experience \cite{Djorgovski2006}. The classification must be updated dynamically as more data come in from the telescope and the feedback arrives from the follow-up facilities. Another problem is determining what follow-up observations are the most useful for improving classification accuracy, and detecting objects of scientific interest. In \cite{Djorgovski2011} maximizing the conditional mutual information is proposed.

	Tackling the future challenges in astronomy will require the cooperation of scientists working in the fields of astronomy, statistics, informatics and machine learning/computational intelligence \cite{Feilgelson2012}. In fact the fields of astroinformatics \cite{Borne2010} and astrostatistics have been recently created to deal with the challenges mentioned above.  Astroinformatics is the new data-oriented paradigm for astronomy research and education, which includes data organization, data description, taxonomies, data mining, knowledge discovery, machine learning, visualization and statistics \cite{Borne2013}.

	The characterization (unsupervised learning) and classification (supervised learning) of massive datasets are identified as major research challenges \cite{Borne2013}.  For time-domain astronomy the rapid detection, characterization and analysis of interesting phenomena and emergent behavior in high-rate data streams are critical aspects of the science \cite{Borne2013}. Unsupervised learning and semi-supervised learning are believed to play a key role in new discoveries.
	To deal with big data in TDA in the Peta-scale era the following open problems need to be solved:

		\noindent \textbf{1) Developing very efficient algorithms for large-scale astroinformatics/astrostatistics}. Fast algorithms for commonly used operations in astronomy are described in \cite{March2012}: \emph{e.g.}, all nearest neighbors, $n$-point correlation, Euclidean minimum spanning tree, kernel density estimation (KDE), kernel regression, and kernel discriminant analysis. $N$-point correlations are used to compare the spatial structure of two data sets, \emph{e.g.}, luminous red galaxies in the Sloan digital sky survey \cite{Kulkarni2007}. KDE is used for comparing the distributions of different kinds of objects. Most of these algorithms involve distance comparisons between all data pairs, and therefore are naively ${\cal O}(N^2)$ or of even higher complexity. With the goal of achieving linear or ${\cal O}(N\log N)$ runtimes for pair-distance problems, space-partitioning tree data structures such as $kd$-trees are used, in a divide and conquer approach. In the KDE problem series expansions for sums of kernels functions are truncated to approximate continuous functions of distance. In \cite{Bengio2007} it is argued that the algorithms should be efficient in three respects: computational (number of computations done), statistical (number of samples required for good generalization), and human involvement (amount of human labor to tailor the algorithm to a task). The authors state that there are fundamental limitations for certain classes of learning algorithms, \emph{e.g.}, kernel methods. These limitations come from their shallow structure (single layered) which can be very inefficient in representing certain types of functions, and from using local estimators which suffer the curse of dimensionality. Contrarily, deep architectures, which are compositions of many layers of adaptive nonlinear components, \emph{e.g.}, multilayer neural networks with several hidden layers, have the potential to generalize in nonlocal ways. In \cite{Hinton2006} a layer-by-layer unsupervised learning algorithm for deep structures was proposed, opening a new line of research that is still on-going.
		
		\noindent \textbf{2) Developing effective statistical tools for dealing with big data}. The large data sample and high dimensionality characteristics of big data, raise three statistical challenges \cite{Fan2013}: i) noise accumulation, spurious correlations, and incidental endogeneity (residual noise is correlated with the predictors), ii) heavy computational cost and algorithmic instability, iii) heterogeneity, statistical biases. Dimension reduction and variable selection are key for analyzing high dimensional data \cite{Fan2013,Donalek2013,Vergara2014}. Noise accumulation can be reduced by using the sparsity assumption.
		
		\noindent \textbf{3) Creating implementations targeted for High-Performance Computing (HPC) architectures}. Traditional analysis methods used in astronomy do not scale to peta-scale volumes of data on a single computer \cite{Das2012}. One could rework the algorithms to improve computational efficiency but even this might prove to be insufficient with the new surveys. An alternative is to decompose the problem into several independent sub-problems. Computations can then proceed in parallel over a shared memory cluster, a distributed memory cluster, or a combination of both. In a shared memory cluster the processes launched by the user can communicate and share data and results through memory. In a distributed environment each processor receives data and instructions, performs the computations and reports the results back to the main server. The number of processors per node, amount of shared memory and network speed have to be taken into account when implementing an algorithm for HPC architectures. Efficiency will ultimately depend on how separable the problem is in the first place. Another parallel computing strategy involves the use of GPUs (graphical processing units) instead or side by side with conventional CPUs (central processing units). GPGPU (general purpose computing in GPU) is a relatively new paradigm for highly parallel applications in which high-complexity calculations are offloaded to the GPU (coprocessor). GPUs are inherently parallel harnessing up to 2,500 processing cores\footnote{NVIDIA Tesla K20 module.}. The processing power and relatively low cost of GPUs have made them popular in the HPC community and their availability has been on the rise \cite{Kindratenko2011}. Note that explicit thread and data parallelism must be exploited in order to get the theoretical speed-ups of GPUs over CPUs. 
		Dedicated hardware based on FPGAs may provide interesting speed-ups for TDA algorithms \cite{Sart2010}. However due to the advanced technical knowledge required to use them, FPGAs are not as popular in astronomy as the HPC resources already presented. Interdisciplinary collaborations between electrical \& computer engineers and astronomers might change this in the near future. 
		An existing non-parallel algorithm can be extended using the MapReduce \cite{Dean2004} model for distributed computing, a model inspired from functional programming. Programs written in this functional style are automatically parallelized. In \cite{Chu2006} the MapReduce model was used to develop distributed and massively parallel implementations of $k$-means, support vector machines, neural networks, Na\"ive Bayes, among others. 
		In the cloud computing paradigm the hardware resources (processors, memory and storage) are almost entirely abstracted and can be increased/decreased by the user on demand. Distributed models such as MapReduce have a high synergy with the cloud computing paradigm. Cloud computing services such as Amazon EC2 provide cost-effective HPC solutions for compute and/or memory bound scientific applications as shown in \cite{Berriman2010}. As pointed out in \cite{Wiley2011}, one of the biggest advantages of implementing astronomical pipelines using cloud services is that the computing resources can be scaled rather easily and according to changing workloads. 
		The granular computing paradigm \cite{Pedrycz2008} is also of interest for astronomical big data applications. In granular computing the information extracted from data is modeled as a hierarchical structure across different levels of detail or scales. This can help to compress the information and reduce the dimensionality of the problem. Another approach is the virtual observatory (VO), a cyberinfrastucture for discovery and access to all distributed astronomical databases \cite{Borne2013}. A useful data portal for data mining is OpenSkyQuery \cite{Greene2007}, which allows users to do multi-database queries on many astronomical object catalogs. 
			
		\noindent \textbf{4) Developing fast algorithms for online event detection and discrimination}. Several facilities around the world will follow the 2 million events that the LSST will issue each night. These facilities will need to decide which events are most relevant so as not to waste their limited observing and storage resources. In addition, these decisions have to be made as fast as possible to avoid missing important data. Pattern recognition methods to quickly analyze and discriminate interesting phenomena from the streamed data are needed. These methods should update their results online and return an associated statistical confidence that increases as more data is retrieved from the LSST. It is critical not to miss any relevant event while keeping the contamination from false positives as low as possible. Additionally, these methods should learn from past experience and adapt depending on the previously selected events. Designing methods that comply with these requirements is currently an open problem.

\section{Concluding Remarks} \label{conclusion}

	In a few years the LSST will be fully operational capturing the light of billions of astronomical objects, and generating approximately two million events each night for ten years. The LSST team itself, and multiple external facilities around the world, will follow and study these events. The main objectives are to characterize and classify the transient phenomena arising from the moving sky. Additionally, it is expected that a plethora of scientific discoveries will be made. If the right tools are used, science would be produced at rates without precedent.

	Conventional astronomy is not prepared for this deluge of observational data and hence a paradigm shift in TDA has been observed. Astronomy, statistics and machine learning have been combined in order to produce science that can provide automated methods to deal with the soon to come synoptic surveys. Computational intelligence methods for pattern recognition are essential for the proper exploitation of synoptic surveys, being able to detect and characterize events that otherwise might not even be noticed by human investigators. 
	
	In this review we have studied several machine learning based implementations proposed to solve current astronomical problems. The particular challenges faced when applying machine learning methods in TDA include 1) the design of representative training sets, 2) the combination and reuse of training databases for new surveys, 3) the definition of feature vectors from domain knowledge, 4) the design of fast and scalable computational implementations of the methods in order to process the TDA databases within feasible times, and finally, 5) the sometimes difficult interpretation of the results obtained and the question of how to gain physical insight from them.

	The quality of a training set is critical for the correct performance of supervised methods. In astronomy an intrinsic sample selection bias occurs when knowledge gathered from previous surveys is used with new data. Semi-supervised learning and active learning rise as feasible options to cope with large and heterogeneous astronomical data, providing particular solutions to the dilemmas regarding training sets. It is very likely that we will see more semi-supervised applications for astronomy in the near future. The reuse of training sets is critical in terms of scalability and validity of results. The integration with existing databases and the incorporation of data observed at different wavelengths are currently open issues. Feature spaces that are survey-independent may provide an indirect solution to the combination of training sets and the applications of trained classifiers across different surveys.

	Although powerful, the sometimes extended calibration required by machine learning methods can be difficult for inexperienced users. The selection of the algorithms, the complexity of the implementations, the exploration of parameter space, and the interpretation of the outputs in physical terms are some of the issues one has to face when using machine learning methods. The learning curve might be too steep for an astronomer to take the initiative, but all the issues named here can be solved by inter-disciplinary collaboration. Teams assembled from the fields of astronomy, statistics, computer science and engineering have everything that is needed to propose solutions for data-intensive TDA. The deluge of astronomical data opens up huge opportunities for professionals with knowledge in computational intelligence and machine learning.

\section{Acknowledgement}
	
	This work was funded by CONICYT-CHILE under grant FONDECYT 1110701 and 1140816, and its Doctorate Scholarship program. Pablo Est\'evez acknowledges support from the Ministry of Economy, Development, and Tourism's Millennium Science Initiative through grant IC12009, awarded to The Millennium Institute of Astrophysics, MAS.


\bibliographystyle{hieeetr}
\addcontentsline{toc}{chapter}{Bibliography}
\bibliography{references}
\end{document}